\documentclass[UKenglish]{lipics}
\pdfoutput=1
\usepackage{authblk}

\usepackage{ifthen}
\usepackage{graphics,epsfig}
\usepackage{amsmath,amsfonts,amssymb}
\usepackage{algpseudocode}
\usepackage{color}
\usepackage{algorithm}

\newcommand{\CONGEST}{{\sc congest}}
\newcommand{\CONGESTB}{{\sc congest}$_B$}
\newcommand{\CONGESTlog}{{\sc congest}$_{O(\log n)}$}
\newcommand{\LOCAL}{{\sc local}}

\title{How Bandwidth Affects the $CONGEST$ Model}

\author[1,2]{Dennis Olivetti\thanks{Additional support from ANR Project DESCARTES.}}

\affil[1]{Gran Sasso Science Institute, L'Aquila, Italy.}
\affil[2]{ Institut de Recherche en Informatique Fondamentale\\CNRS and University Paris Diderot, France.}
\authorrunning{Dennis Olivetti}

\date{}
\begin{document}

\maketitle

\begin{abstract}	
	The \CONGEST\/ model for distributed network computing is well suited for analyzing the impact of limiting the throughput of a network on its capacity to solve tasks efficiently. For many ``global'' problems there exists a lower bound of $\Omega(D + \sqrt{n/B})$, where $B$ is the amount of bits that can be exchanged between two nodes in one round of communication, $n$ is the number of nodes and $D$ is the diameter of the graph. Typically, upper bounds are given only for the case $B=O(\log n)$, or for the case $B = +\infty$. 

	In this work we study tradeoffs between the round complexity for solving a task, and the bandwidth of the links. We show three different classes of problems, namely, \textit{bandwidth efficient}, \textit{bandwidth sensitive} and \textit{bandwidth insensitive}. The first is the class of problems having a round complexity that fully scales with the bandwidth. We show that the All Pairs Shortest Path problem belongs to this class. The second class includes all problems having a round complexity that scales with the bandwidth, but not linearly. This is the case of the Minimum Spanning Tree construction and the Single Source Shortest Path problems. At last, the third class contains problems whose complexity does not depend on the size of the messages. We show that this class is not empty.
\end{abstract}

\section{Introduction}	
\subsection{Context}
The links of networks typically have limited bandwidth. The \CONGEST\/ model for distributed network computing captures this constraint. In this model, an algorithm proceeds in synchronous rounds. At each round, every node can send $B$ bits to each of its neighbors in the network (these $B$ bits do not need to be the same for all neighbors). A typical value for $B$ is $ O(\log n)$ in $n$-node networks. This value is sufficient to transmit an integer in a polynomial range, like the identifier of a node, or the weight of an edge.

One celebrated result in this context is a minimum-weight spanning tree (MST) construction algorithm that performs in $O(D + \sqrt{n} \log^*n)$ rounds in diameter-$D$ $n$-node networks \cite{DBLP:journals/jal/KuttenP98}. This complexity is optimal for $B = O(\log n)$ \cite{DBLP:conf/sofsem/OokawaI15}. On the other hand, all (computable) tasks can be solved in $O(D)$ rounds in diameter-$D$ networks whenever there is no limitation on the bandwidth, by gathering all data at one node, computing the solution at that node, and broadcasting that solution to all nodes.

The aim of this work is to investigate tradeoffs between the round complexity for solving a task, and the bandwidth of the links. So far, results are known only for $B=O(\log n)$ in the classical \CONGEST\/ model, and $B=+\infty$ in the so-called \LOCAL\/ model. A specific case that deserves particular interest is to determine, given a task, the minimum value for B such that the task is solvable in $O(D)$ rounds in diameter-$D$ networks with links of bandwidth $B$. For instance, in the case of the MST construction task, what is the minimum value of $B$ such that MST can be constructed in $O(D)$ rounds? It is known that $B = O(\log n)$ is not sufficient, as $\Omega(D + \sqrt{n})$ rounds is a lower bounds on the number of rounds for such a value of $B$ \cite{DBLP:conf/sofsem/OokawaI15}.

In this paper, our objective is twofold. First, we are aiming at establishing tradeoffs between the size of the messages $B$ and the number of rounds, by obtaining algorithms for the \CONGESTB\/ model, having round complexities that are parametric on $B$. Our second objective is to better understand the role of $B$ in the \CONGESTB\/ model, and try to figure out if $B$ always has some impact on the round complexity.

\subsection{Our Results}
Let $T_P(\mathcal{X})$ be the exact round complexity of a problem $P$ using messages of size $B= O(\mathcal{X} \log n )$, given by ignoring additive factors that do not depend on $B$ (e.g., the diameter). We define the speedup as $S_P(\mathcal{X}) = \frac{T_P(1)}{T_P(\mathcal{X})}$. Depending on $S_P$, we define three classes of problems and, for each of these classes, we provide examples of problems belonging to them.
\begin{itemize}
	\item \textit{bandwidth efficient}: the class of all problems $P$ such that $S_P(\mathcal{X}) = \Theta(\mathcal{X})$, i.e., problems having a round complexity that fully scales with the bandwidth.
	\item \textit{bandwidth sensitive}: this class contains all problems $P$ such that $S_P(\mathcal{X}) = o(\mathcal{X})$ and $S_P(\mathcal{X})=\Omega(1)$, i.e., those problems that have a round complexity that scales with the bandwidth, but not linearly.
	\item \textit{bandwidth insensitive}: the class of all problems $P$ such that $S_P(\mathcal{X}) = \Theta(1)$, i.e., whose complexity does not depend on the size of the messages. 
\end{itemize}

First, we investigate the round complexity, in the \CONGESTB\/ model, of the All Pairs Shortest Path (APSP) problem in unweighted graphs. We know that for this problem there is a lower bound of $\Omega(D+ \frac{n}{B})$ rounds \cite{DBLP:conf/soda/FrischknechtHW12} and an algorithm performing in $O(n)$ rounds \cite{DBLP:conf/podc/HolzerW12}. We show that the APSP problem is bandwidth efficient, i.e., that $S_{APSP}(\mathcal{X}) = \Theta(\mathcal{X})$, by modifying the existing algorithm to run in time $O(D+ \frac{n \log n}{B}) = O(D+ \frac{n}{\mathcal{X}})$.

We then investigate the round complexity of two well studied problems, namely Minimum Spanning Tree (MST) and Single Source Shortest Path (SSSP). For both these problems there is a lower bound of $\Omega(D + \sqrt{\frac{n}{B}})$ rounds \cite{SarmaHKKNPPW11}. On the other hand, these two problems have been studied only when $B=O(\log n)$. In this case, there exists an algorithm that solves the MST construction problem in $O(D + \sqrt{n}\log^*n)$ rounds \cite{DBLP:journals/jal/KuttenP98}, that is a round complexity that nearly matches the lower bound. For the SSSP problem, no sublinear algorithm that matches the lower bound and finds an exact solution is known. Conversely, we can find a $(1+\epsilon)$-approximation of the solution in $\widetilde{O}(\epsilon^{-O(1)} (\sqrt{n} + D) )$ rounds, that is optimal. We show that the round complexity of both these problems scales with $B$, that is, if it is possible to send messages of size $B$ at each round,
\begin{itemize}
	\item there exists an algorithm that constructs a MST in $\widetilde{O}(D + \sqrt{\frac{n}{B}})$ rounds;
	\item there esists an algorithm that finds a $(1+\epsilon)$-approximation of SSSP in $\widetilde{O}(\epsilon^{-O(1)} (\sqrt{\frac{n}{B}} + D) )$ rounds.
\end{itemize}
Notice that these two round complexities match their respective lower bounds, for any value of $B$.  Also, for both these problems, $S(\mathcal{X}) = \Theta(\sqrt{\mathcal{X}})$, i.e., these two problems are \textit{bandwidth sensitive}.

We then show that there are problems for which, by increasing the bandwidth of the links, the round complexity does not change. In order to reduce the round complexity, one needs to increase the size of the messages exponentially. More specifically, we show that there is a problem, $Distance_k$, solvable in $O(\log n)$ rounds using messages of size $B=O(\log n)$, but such that if one wants to solve that problem in $\log n - 1$ rounds, then one needs to use messages of size at least $\Omega( \frac{n}{\log^3 n} )$. In other words, $Distance_k$ is a \textit{bandwidth insensitive} problem for bandwidths in the range from $\Omega(\log n)$ to $O(\frac{n}{\log^3 n})$ bits, since in this range $S(\mathcal{X}) = \Theta(1)$.

\section{Related Work}
In general, the MST construction problem has been widely studied. In the distributed asynchronous context, Gallager, Humblet and Spira~\cite{DBLP:journals/toplas/GallagerHS83} gave an algorithm with time complexity $O(n \log n)$ that uses $O(|E| + n \log n)$ messages, which is optimal. In the synchronous setting, the first sublinear algorithm was given by Garay et al. in \cite{GarayKP98}. Its running time is $O(D + n^{\frac{\ln 3}{\ln 6}} \log^* n) \approx O(D + n^{0.61} \log^* n)$, where $D$ is the diameter of the graph. This complexity was later improved to $O(D + \sqrt{n} \log^* n)$ in \cite{DBLP:journals/jal/KuttenP98}. Then, Peleg et al.~\cite{PelegR00} showed that this latter complexity is nearly optimal, giving an $\Omega(D + \frac{\sqrt{n}}{\log n})$ lower bound, which was improved by Sarma et al.~\cite{SarmaHKKNPPW11} to $\Omega(D + \sqrt{\frac{n}{\log n}})$ and then by Ookawa et al.~\cite{DBLP:conf/sofsem/OokawaI15} to $\Omega(D + \sqrt{n})$. All these lower bounds hold for graphs with diameter $\Omega(\log n)$. For constant diameter graphs, there are bounds of $\widetilde{\Omega}(n^{1/3})$ rounds for diameter~4, $\widetilde{\Omega}(n^{1/4})$ rounds for diameter~3, and $O(\log n)$ rounds for diameter~2 (see \cite{LotkerPP01}). Then, Elkin~\cite{Elkin04} showed that, if termination detection is not required, the diameter of the graph is not a lower bound, and that there exists an algorithm that requires $\widetilde{O}(\mu + \sqrt{n})$ rounds, where $\mu$ is the so-called MST-radius of the graph. Recently, Pandurangan et al. \cite{DBLP:journals/corr/Pandurangan0S16a} showed a randomized algorithm that is able to construct a MST in $\widetilde{O}(D + \sqrt{n})$ rounds while using an optimal number of messages, $\widetilde{O}(|E|)$. Elkin \cite{DBLP:journals/corr/Elkin17} showed how to achieve the same result deterministically. Notice that all the aforementioned algorithms use messages of size $B=O(\log n)$, while some of the lower bounds have been explicitly stated as a function of $B$. Among the various lower bounds, the best one that depends on $B$ is the one of Sarma et al.~\cite{SarmaHKKNPPW11}, that is $\Omega(D + \sqrt{\frac{n}{B}})$. In the congested clique the MST problem can be solved much faster: Ghaffari et al. \cite{DBLP:conf/podc/GhaffariP16} gave an algorithm that solves the problem in $O(\log^* n)$ rounds using messages of size $O(\log n)$. They also showed that, by using slightly bigger messages, the complexity becomes $O(1)$ rounds.

Concerning the Single Source Shortest Path problem, an exact solution can be found using Bellman \& Ford Algorithm in linear time. The problem of finding an exact solution in sublinear time remained opened for years, and only very recently Elkin \cite{DBLP:journals/corr/Elkin172} gave an algorithm that is able to solve the problem in $O((n \log n)^\frac{5}{6})$ rounds for $D=O(\sqrt{n \log n})$ and $O(D^\frac{1}{3} \cdot (n \log n)^\frac{2}{3})$ rounds for larger diameters. The problem of finding an approximate solution seems easier. In fact, Lenzen et al. \cite{DBLP:conf/stoc/LenzenP13} showed how to find an $O(\frac{1}{\epsilon})$-approximation in $\widetilde{O}(n^{\frac{1}{2} + \epsilon} +D)$ rounds. Then, Nanongkai \cite{DBLP:conf/stoc/Nanongkai14} improved this result providing a randomized algorithm that finds a $(1+\epsilon)$-approximation in $\widetilde{O}(n^\frac{1}{2}\cdot D^\frac{1}{4} + D)$ rounds. Finally, Henzinger et al. \cite{DBLP:conf/stoc/HenzingerKN16} removed the multiplicative dependence on $D$, by giving a deterministic algorithm that solves the problem in $O(n^{\frac{1}{2} + o(1)} + D^{1+o(1)} )$ and finds a $(1+o(1))-$approximation. Becker et al. \cite{DBLP:journals/corr/BeckerKKL16} improved this complexity by showing how to find a $(1+\epsilon)$ approximation in $\widetilde{O}(\epsilon^{-O(1)} \cdot (n^\frac{1}{2} + D))$ rounds.
This last algorithm matches the lower bound given by Sarma et al.~\cite{SarmaHKKNPPW11}, that is $\Omega(D + \sqrt{\frac{n}{B}})$, but all the aforementioned algorithms are for $B=O(\log n)$.

The All Pairs Shortest Path problem is also well studied. Frischknecht et al.~\cite{DBLP:conf/soda/FrischknechtHW12} showed that, in dense graphs, the diameter can not be computed in sublinear time (using small messages), by providing a lower bound of $\Omega(\frac{n}{B})$ rounds. This result implies a lower bound for the APSP problem as well. Then, Abboud et al.~\cite{DBLP:conf/wdag/AbboudCK16} provided an $\widetilde{\Omega}(n)$ lower bound for $B=O(\log n)$, even for sparse networks. Concerning upper bounds, Holzer et al. \cite{DBLP:conf/podc/HolzerW12} showed how to solve the APSP problem deterministically, in $O(n)$ rounds, in unweighted graphs. Nanongkai~\cite{DBLP:conf/stoc/Nanongkai14} presented a randomized algorithm that finds a $(1+o(1))$-approximation in $\widetilde{O}(n)$ rounds in the weighted case. Lenzen et al. \cite{DBLP:conf/podc/LenzenP15} showed that a $(1+\epsilon)$-approximate solution can be deterministically found in $O(\epsilon^{-2} \cdot n \log n)$ rounds. 

\section{Model and Definitions}

We are considering the classical \CONGEST\/ model for distributed network computing~\cite{Peleg2010}. The network is modeled as a connected simple graph (no self-loops, and no parallel edges). The nodes of the graph are computing entities exchanging messages along the edges of the graph. Nodes are given arbitrary distinct identities (IDs) in a range polynomial in~$n$, in $n$-node networks. Hence, every ID can be stored on $O(\log n)$ bits. 

In the \CONGEST\/  model, all nodes start simultaneously, and execute the same algorithm in parallel. Computation proceeds synchronously, in a sequence of \emph{rounds}. At each round, every node:
\begin{itemize}
	\item performs some individual computation,
	\item sends messages to neighbors in the network, and 
	\item receives messages sent by neighbors. 
\end{itemize}
The main constraint imposed by the \CONGEST\/ model is a restriction of the amount of data that can be transferred between neighboring nodes during a round. The model in which the messages are limited to $B$ bits is called \CONGESTB\/. Typically $B$ is chosen to be $O(\log n)$, and \CONGEST\/ is typically used to refer to the \CONGESTlog\/ model.

The \CONGEST\/ model is well suited for analyzing the impact of limiting the throughput of a network on its capacity to solve tasks efficiently. The complexity of a distributed algorithm in the \CONGEST\/ model is expressed in number of rounds.

A similar model is the Congested Clique, where the communication graph is a clique and the same restrictions of the \CONGEST\/ model hold, i.e., all nodes can communicate with all the other nodes in one round using messages of size $B$. In the Broadcast Congested Clique each node is restricted to send the \textit{same} message to all the other nodes.

\section{All Pairs Shortest Path}
In this section we show how to modify the APSP algorithm of \cite{DBLP:conf/podc/HolzerW12} and reduce the round complexity when the bandwidth increases. Recall that, in the APSP problem, each node needs to find its distance from all the other nodes. In \cite{DBLP:conf/podc/HolzerW12} is shown how to deterministically find an exact solution of the APSP problem in $O(D + n)$ rounds in unweighted graphs. The procedure is the following:
\begin{enumerate}
	\item construct a BFS tree;
	\item perform a DFS visit on the tree;
	\item at each step of the DFS visit, wait 1 round and then start a BFS from the current node (if it is visited for the first time).
\end{enumerate}
All nodes will know their distances from all the other nodes by knowing their distance from the root of each BFS tree. In \cite{DBLP:conf/podc/HolzerW12} is shown that, by waiting one round before starting each visit, the breadth-first searches can be executed concurrently without congestion. In other words, at each round, at most one message passes on a fixed edge. The complexity comes by the fact that the DFS visit requires $O(n)$ rounds and that $O(D)$ is the time required to complete a BFS.

We show how to modify this procedure in order to solve the problem in $O(D + \frac{n \log n}{B})$ rounds.

\begin{theorem}\label{thm:apsp}
	There exists an algorithm for the \CONGESTB\/ model that solves the All Pairs Shortest Path problem in $O(D + \frac{n \log n}{B})$ rounds.
\end{theorem}

\begin{proof}
	Let $s_i$, $1\le i\le n$, be the $i$-th node visited (for the first time) by a DFS performed on the BFS tree. Let $t_i$, $0\le t_i < 2n$, be the time at which node $s_i$ is visited \emph{for the first time} by the DFS. Consider the sequence $\mathcal{T} = (t_1,\ldots,t_n)$. We split $\mathcal{T}$ in $\lceil \frac{B}{\log n} \rceil$ subsequences $T_1, \ldots, T_{\lceil\frac{B}{\log n}\rceil}$ of length at most $\lceil \frac{2\,n \log n}{B} \rceil$ each, where $T_j$ contains the values $t_i$ such that $\lceil \frac{2\,n \log n}{B} \rceil (j-1)   \le t_i < \lceil \frac{2\,n \log n}{B} \rceil j$. Notice that, if each node $s_i$ starts its BFS at time $2\, t_i$, there is no congestion between different breadth-first searches, for the same arguments showed in~\cite{DBLP:conf/podc/HolzerW12}. Now, in order to speed up the computation, we can visit all the sequences concurrently, by starting the BFS of node $s_i$ with $t_i \in T_j$ at time $2(t_i - \lceil \frac{2\,n \log n}{B} \rceil (j-1))$. In other words, we are splitting the DFS visit in at most $\lceil \frac{B}{\log n} \rceil$ parts, each of them of length at most  $\lceil \frac{2\,n \log n}{B} \rceil$. By doing this, there can be congestion for at most $\lceil \frac{B}{\log n} \rceil$ breadth-first searches at the same time (one for each sequence), and since the bandwidth is $B$, a constant number of rounds is enough to transmit on the same edge $\lceil \frac{B}{\log n} \rceil $ messages of size $O(\log n)$ belonging to different breadth-first searches. Since each sequence is of length at most $\lceil \frac{2\,n \log n}{B}\rceil$, the total time required to solve the problem is $O( D + \frac{n \log n}{B} )$.

	What remains to show is how each node $s_i$ can compute $t_i$ in $O(D)$ rounds. Notice that, by performing a convergecast, all nodes of the BFS tree can learn the size of the subtree rooted at each of its children. Now, starting from the root $s_1$ and setting $t_1 = 0$, each node $s_i$ can assign an arbitrary order to its children, and send, to each child $s_j$, $t_i$ and the sum $\sigma$ of the number of nodes present in the subtrees rooted on the children before $s_j$. Then each child $s_j$ can compute $t_j = 2 \sigma + t_i + 1$ and repeat the same procedure. In total, $O(D)$ rounds are required.

\end{proof}

\section{Minimum Spanning Tree}
In this section we show how to modify the MST construction algorithm of \cite{DBLP:journals/jal/KuttenP98} and reduce the round complexity when the bandwidth increases, by proving the following theorem.
\begin{theorem}\label{thm:mst}
	There exists an algorithm for the \CONGESTB\/ model that constructs a Minimum Spanning Tree in $\widetilde{O}(D + \sqrt{\frac{n}{B}})$ rounds.
\end{theorem}

Recall that, for the MST construction task, every node is given as input the weight $w(e)$ of each of its incident edges $e$. These weights are supposed to be of values polynomial in the size $n$ of the network $G=(V,E,w)$, and thus each weight can be stored on $O(\log n)$ bits. The output of every node is a set of incident edges, such that the collection of all outputs forms an MST of the network. A Minimum Spanning Tree is a subset of edges $T \subseteq E$ such that $(V,T)$ is connected and $\sum_{e \in T} w(e)$ is minimum.  At the end of the computation each node must know which of its adjacent edges belong to the Minimum Spanning Tree. 

The algorithm of \cite{DBLP:journals/jal/KuttenP98} is divided in two phases: {\sc Fast-Dom-G} and {\sc Pipeline}. The first part, {\sc Fast-Dom-G}, computes a set of sub-MST by applying a modified version of the algorithm of Gallager, Humblet and Spira (GHS) \cite{DBLP:journals/toplas/GallagerHS83}. The second part, {\sc Pipeline}, finds the remaining edges by performing a convergecast, where nodes send edges in non-decreasing order of weight by ignoring edges that close some cycle. {\sc Fast-Dom-G} runs in $O(k \log^* n)$ rounds and creates $\frac{n}{k+1}$ fragments, where $k$ is a parameter to be fixed later, while {\sc Pipeline} runs in $O(D + \frac{n}{k})$ rounds. By setting $k = \sqrt{n} $, they obtain a complexity of $O(D+\sqrt{n} \log^* n)$ rounds.

We show that {\sc Pipeline} can be performed faster if more bandwidth is allowed. More specifically, we show that it can be performed in $O(D+\frac{n \log n}{k \cdot B})$ rounds. By choosing $k = \sqrt{\frac{n \log n}{B}}$, we obtain Theorem \ref{thm:mst}.

\begin{lemma}
	\label{fastpart2}
	{\sc Pipeline} can be completed in $O(D+\frac{n \log n}{k \cdot B})$ rounds, where $B$ is the size of the messages.
\end{lemma}

The rest of the section is used to prove Lemma \ref{fastpart2}, that is, we now show how to adapt the {\sc Pipeline} procedure in such a way that if each node sends $O(\frac{B}{\log n})$ edges per round, its running is still fully pipelined. This implies that it is possible to send $O(\frac{n}{k+1})$ edges in $O(D+\frac{n}{k+1} / \frac{B}{\log n}) = O(D+\frac{n \log n}{(k+1) B}) = O(D + k)$ rounds.

The existing {\sc Pipeline} algorithm, modified to send $\frac{B}{\log n}$ edges per round, does the following:
\begin{enumerate}
	\item construct a BFS-tree of G;
	\item each node knows, from the first phase ({\sc Fast-Dom-G}), the edges that connect it to other fragments;
	\item each node keeps a set of edges (in non-decreasing order of weight) not already sent to its parent;
	\item leaves start at round 0, intermediate nodes start when they have received at least a message from all their neighbors;
	\item at each step each node sends the $\frac{B}{\log n}$ lightest edges that does not close a cycle with the edges already sent (or the ones that it is sending) and repeats this step until there are no valid edges (since $\frac{B}{\log n}$ could be non-integral, we can use two rounds to complete this task);
	\item at the end the root chooses the $\frac{n}{k+1}$ lightest received edges and broadcasts them.
\end{enumerate}

We adapt the proof in \cite{DBLP:journals/jal/KuttenP98} to show that the {\sc Pipeline} procedure described above correctly completes within the required time. We first show the main part of the proof of \cite{DBLP:journals/jal/KuttenP98} and then our adaptation. 

In \cite{DBLP:journals/jal/KuttenP98} is shown inductively that, at each step, there is at least one edge that can be upcasted. Let a node be \textit{active} if it has sent an edge in the previous step. Assume that there is a node $v$ that is active from $m$ steps and it has at least one active child $u$. Notice that its active children are active from at least $m+1$ steps. Let $U$ be the set of edges already sent from node $v$ to its parent in the previous steps. We know that this set forms a forest of trees, that we call $U_1, ... , U_l$. Let $x_i = |U_i|$. Notice that $\sum x_i = m$, since node $v$ sent an edge in all the previous $m$ rounds. We also know that $|V(U_i)| = x_i +1$, since $U_i$ is a tree. Let $C$ be the set of edges that $v$ already received from its child $u$. We know that $|C| \ge m+1$, since each child is active from at least one step before its parent.

Now, suppose for the sake of contradiction that all the edges in $C$ are not candidates that can be sent. It means that each edge, either it has already been sent, or that it closes a cycle. In both cases, the endpoints of each edge are part of some edge of $U$ (the edges already sent). Since $U$ forms a forest this also implies that there are no edges that connect two different trees $U_i, U_j$ of $U$. It means that $C$ can be partitioned depending on the trees to which the endpoints of each edge belong to. We can say that for each $i$, each edge in $C_i$ has its endpoints in $V(C_i)$. And since $C$ is a forest, we obtain that $|C_i| \le |V(U_i)-1|$. It follows that $|C_i| \le |V(U_i)-1| = x_i$, $C \le \sum x_i = m \le |C| -1$ that is a contradiction. This implies that $C$ contains at least one candidate.

The above proof can be adapted for our purposes in the following way. Since at each step a child sends $\frac{B}{\log n}$ edges, we know that $\sum x_i = |U| = \frac{m B}{\log n}$. We know also that children sent edges for at least $m+1$ steps, thus $|C| \ge \frac{(m+1) B}{\log n}$. Again, for the sake of contradiction, assume that the number of candidate edges is less than $\frac{B}{\log n}$. This means that at least $|C| - \frac{B}{\log n} +1$ edges have both their endpoints in $U$. In this case there are at least $|C| - \frac{B}{\log n} +1$ edges that can be partitioned in sets $C_1,\dots , C_l$ such that the endpoints of edges in $C_i$ are only in $V(U_i)$. Hence, we obtain that $\sum |C_i| \le \sum (|V(U_i)| -1 ) = \sum x_i$. Since $\sum |C_i|$ is at least $|C|-\frac{B}{\log n} +1$, we have that $|C| - \frac{B}{\log n} +1 \le \sum x_i$, implying that $|C| \le \sum x_i + \frac{B}{\log n} -1$.
Thus, we obtain $\frac{(m+1) B}{\log n} \le |C| \le \sum x_i + \frac{B}{\log n} -1 = \frac{m B}{\log n} + \frac{B}{\log n} -1 = \frac{(m+1) B}{\log n} -1$, which is a contradiction. This proves that $C$ contains at least $\frac{B}{\log n}$ candidates. 

\begin{figure}[h]
	\centering
	\input{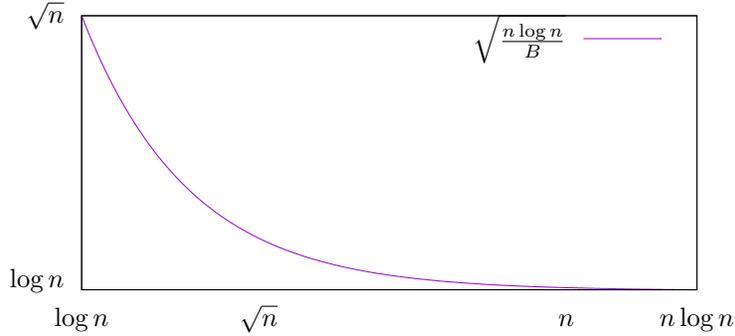}
	\caption{Round complexity as a function of B}
\end{figure}

\section{Single Source Shortest Path}
In this section we show how to modify the SSSP algorithm of \cite{DBLP:journals/corr/BeckerKKL16} and reduce the round complexity when the bandwidth increases. Recall that in the SSSP problem, given a node $v$, all nodes of the graph have to find their distance from $v$. In \cite{DBLP:journals/corr/BeckerKKL16} it is shown how to deterministically find a $(1+\epsilon)$ approximate solution of the SSSP problem using $\widetilde{O}(\epsilon^{-O(1)}(\sqrt{n}+D))$ rounds, where $D$ is the (hop) diameter of the network. First, it exploits a Theorem of \cite{DBLP:conf/stoc/HenzingerKN16}, that states the following:

\begin{theorem}(\cite{DBLP:conf/stoc/HenzingerKN16})
	\label{thmreduce}
	Given any weighted undirected network $G=(V,E,w)$ and a source node $s \in V$, there is a $\widetilde{O}(\sqrt{n})$-round deterministic distributed algorithm in the broadcast \CONGEST\/ model that computes an overlay network $G' = (V', E',w')$ with edge weights $w' : E' \rightarrow \{1, \ldots , poly(n)\}$ and some additional information for every node with the following properties:
	\begin{itemize}
		\item $|V'| = \widetilde{O}(\epsilon^{-1} \sqrt{n})$ and $s \in V'$.
		\item For $\epsilon' := \epsilon/7$, each node $v \in V$ can infer a $(1 + \epsilon)$-approximation of its distance to s from $(1 + \epsilon')$-approximations of the distances between $s$ and each $t \in V'$ .
	\end{itemize}
\end{theorem}

By applying Theorem \ref{thmreduce}, the SSSP problem is reduced to a graph of roughly $\sqrt{n}$ nodes (\emph{skeleton} nodes). Then, in \cite{DBLP:conf/stoc/HenzingerKN16} authors provide an algorithm that solves the SSSP problem in the Broadcast Congested Clique and that runs in $\epsilon^{-9}  polylog(n)$ rounds. They also show how to emulate this algorithm in the \CONGEST\/ model, with a slowdown proportional to $D$ and to the number of nodes of the clique. Since the problem is reduced to an instance with $\sqrt{n}$ nodes, the algorithm can be emulated in $\epsilon^{-9} polylog(n) \cdot O(D + \sqrt{n})$ rounds. At this point the skeleton nodes can broadcast their distance from $s$ to all the nodes of the original graph in $O(D + \sqrt{n})$ rounds, and all the nodes can compute their approximate distance from $s$ (Theorem \ref{thmreduce}).

As stated in \cite{DBLP:conf/stoc/HenzingerKN16}, Theorem \ref{thmreduce} is a deterministic version of a more general Theorem stated in \cite{DBLP:conf/stoc/Nanongkai14}, where $|V'|$ can be of size $\widetilde{O}(\alpha)$ and the running time of their algorithm is $\widetilde{O}(\alpha + \frac{n}{\alpha} + D)$ and succeeds with high probability. The round complexity is proved by providing an algorithm that solves the bounded-hop multi-source shortest path problem in $\widetilde{O}(|V'|+h+D)$ rounds, where the sources are nodes in $V'$ and the number of hops $h$ is $ \frac{n \log n}{\alpha}$.

We now show that the aforementioned algorithm can complete in $\widetilde{O}(\frac{\alpha}{B} + \frac{n}{\alpha} + D)$ rounds, and that a round of communication of the broadcast congested clique can be emulated in $\widetilde{O}(\frac{|V'|}{B})$ in the \CONGESTB\/ model, by proving the following lemmas.

\begin{lemma}
	\label{lemmafastsim}
	A round of communication of the Broadcast Congested Clique can be emulated in the \CONGESTB\/ model in $\widetilde{O}(\frac{|V'|}{B})$ rounds.
\end{lemma}

\begin{proof}
	As in \cite{DBLP:journals/corr/BeckerKKL16}, we can solve the problem by using pipelining on a BFS tree. As in the MST case (Lemma \ref{fastpart2}), the speed of the pipelining linearly depends on $\frac{1}{B}$.
\end{proof}

\begin{lemma}
	\label{lemmafastbhmssp}
	There is a distributed algorithm that runs in $\widetilde{O}(\frac{|V'|}{B}+h+D)$ rounds in the \CONGESTB\/ model that solves the bounded-hop multi-source shortest path problem.
\end{lemma}

\begin{proof}
	The idea of the algorithm in \cite{DBLP:conf/stoc/Nanongkai14} is to execute many bounded-hop single-source shortest path algorithms in parallel, one for each source node, and to randomly delay the starting time of each execution in order to avoid congestion. In \cite{DBLP:conf/stoc/Nanongkai14} is shown that the execution of a single bounded-hop single-source shortest path algorithm requires $O(h + D)$ rounds using messages of logarithmic size, and by choosing a random delay from the interval of numbers from $0$ to $|V'| \log n$, $|V'|$ executions can be performed in parallel without much congestion, obtaining a round complexity of $\widetilde{O}(|V'|+ h + D)$. We show that, by increasing the bandwidth (thus allowing more congestion), it is possible to reduce the size of the interval from where the random delay is chosen.
	
	Let $k=|V'|$. Let $\mathcal{M}_{i,u}$ be the set of messages sent by node $u$ while executing the bounded-hop single-source shortest path algorithm for source $s_i$. In \cite{DBLP:conf/stoc/Nanongkai14} (Lemma 3.7) it is shown that $\mathcal{M}_{i,u} \le c \log n$ for some constant $c$. They then show that, if the delay is randomly chosen from the integers from $0$ to $k \log n$, then the probability that there exists a time $t$, a node $u$ and a set $\mathcal{M} \subseteq \bigcup_i \mathcal{M}_{i,u}$ such that $|\mathcal{M}| \ge \log n$ and all messages in $\mathcal{M}$ are broadcasted by $u$ at time $t = O(k + h + D)$, is $O(\frac{1}{n^2})$. We extend this result showing in the remaining part of the section that, if the delay is randomly chosen from the integers from $0$ to $\frac{k \log^2 n}{B}$, then the probability that $\mathcal{M} \ge B$ is also $O(\frac{1}{n^2})$, that implies the Lemma.
	
	Fix any node $u$, time $t$ and set $\mathcal{M}$ as above. Since $u$ at each round sends at most one message for each $s_i$, we can assume that $|\mathcal{M} \cap \mathcal{M}_{i,u}| \le 1$. This implies that $|\mathcal{M}| \le k$. As in the original proof, we can bound the number of possible sets $\mathcal{M}$ of size $m$ by $\binom{k}{m} (c \log n)^m$. This holds since each set $\mathcal{M}$ can be constructed by picking $m$ different sets $\mathcal{M}_{i,u}$ and picking one message out of $c \log n$ messages from each $\mathcal{M}_{i,u}$. Then, the probability that all the messages of a set are sent at the same round is at most $(\frac{B}{k \log^2 n})^{|M|}$. Thus, for fixed $u$ and $t$, the probability that there exists $\mathcal{M}$ such that $|\mathcal{M}| \ge B$ and all messages in $\mathcal{M}$ are sent by $u$ at time $t$, is at most
	$$  \sum_{m=B}^k \binom{k}{m} (c \log n)^m (\frac{B}{k \log^2 n})^m \le $$
	$$  \sum_{m=B}^k (\frac{k e}{m})^m (c \log n)^m (\frac{B}{k \log^2 n})^m = $$
	$$  \sum_{m=B}^k (\frac{ e c B}{m \log n})^m  \le  \sum_{m=B}^k (\frac{ e c B}{B \log n})^B  \le $$
	$$  \sum_{m=B}^k (\frac{ e c}{\log n})^B  \le k (\frac{ e c}{\log n})^B $$
	Since $B \ge \log n$, for large $n$ the above formula is at most $\frac{1}{n^4}$. Then the lemma follows by summing this probability over all nodes $u$ and time steps $t$.
\end{proof}

Now we have all the ingredients to prove the following theorem.

\begin{theorem}
	\label{algossspB}
	There is a distributed algorithm that runs in $\widetilde{O}(\epsilon^{-O(1)}(\sqrt{\frac{n}{B}}+D))$ rounds in the \CONGESTB\/ model that solves the $(1+\epsilon)$ approximate SSSP problem.
\end{theorem}
\begin{proof}
	Let $|V'|$ be $\widetilde{O}(\epsilon^{-1} \sqrt{n B})$. By Theorem \ref{thmreduce} and Lemma \ref{lemmafastbhmssp}, we can reduce the problem to a SSSP instance on $|V'|$ nodes in $\widetilde{O}(\epsilon^{-1} \sqrt{\frac{n}{B}} +D)$ rounds. We can then simulate, using Lemma \ref{lemmafastsim}, the Broadcast Congested Clique algorithm of \cite{DBLP:conf/stoc/HenzingerKN16} in  $\widetilde{O}(\epsilon^{-O(1)} \sqrt{\frac{n}{B}})$ rounds to solve the original problem.
\end{proof}

\section{$Distance_k$}
\begin{figure}[h]
	\centering
		\includegraphics[width=0.4\textwidth]{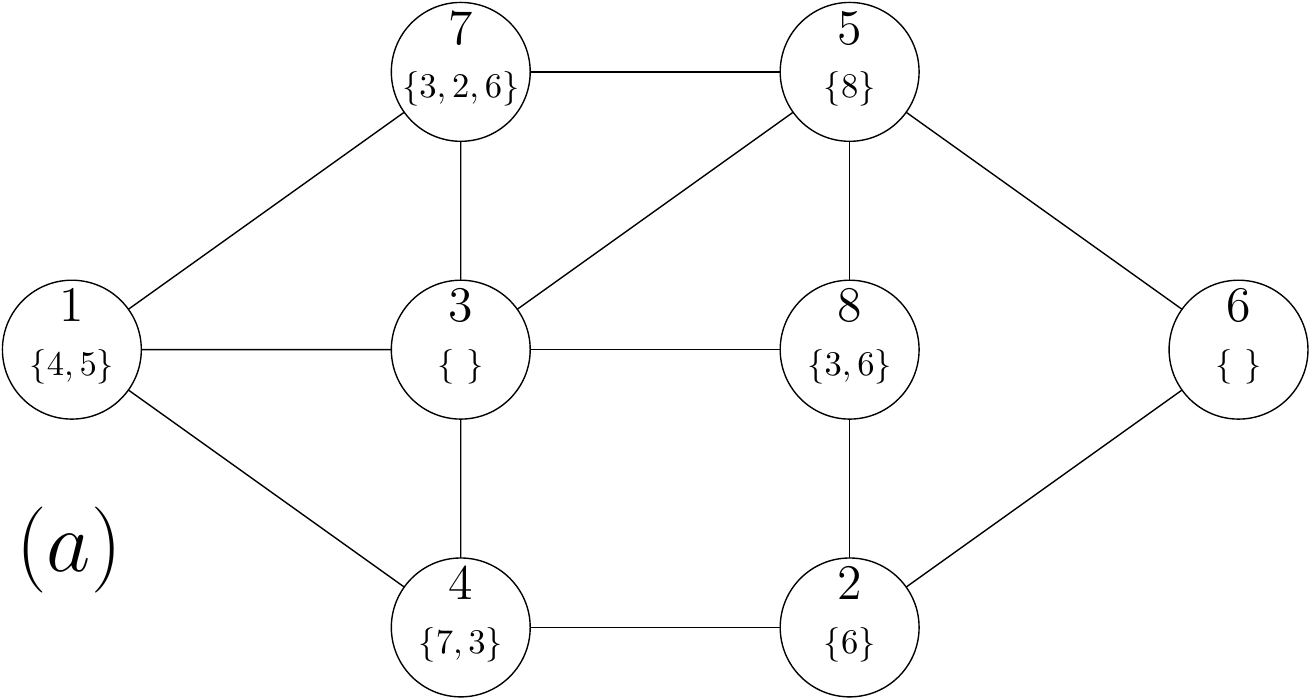}
		\includegraphics[width=0.4\textwidth]{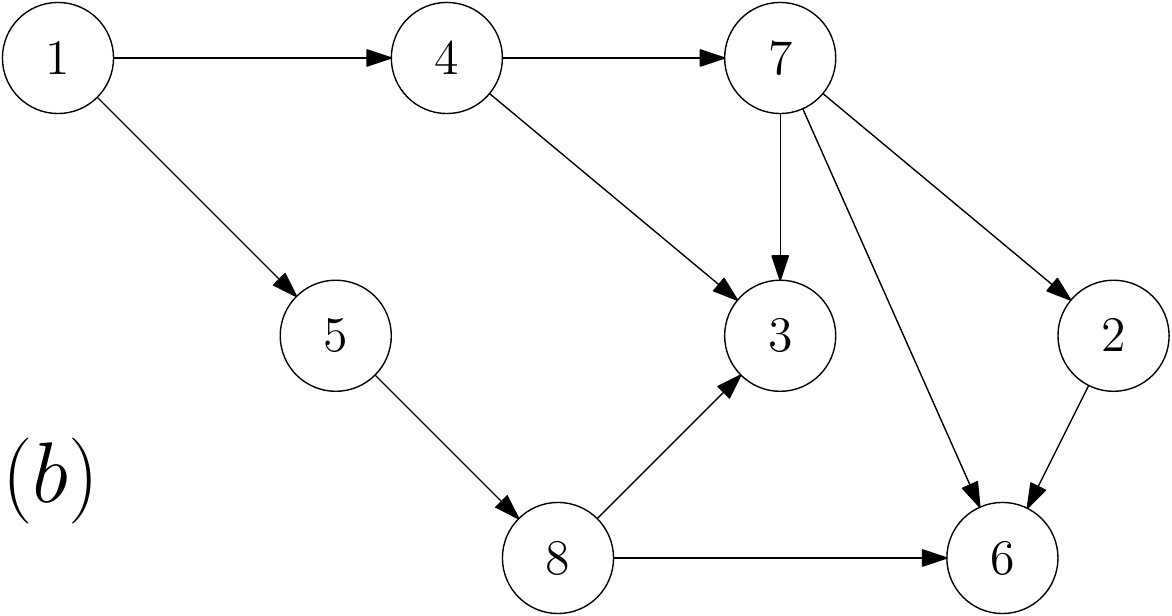}
	\caption{(a) represents $G$, the communication graph where the algorithm is executed and the input given to the nodes, while (b) is $G'$, the directed graph given in input to the nodes.}
	\label{figexampledist}
\end{figure}

In this section we show a problem, called $Distance_k$, whose round complexity does not depend on the size $B$ of the messages. Consider a graph $G=(V,E)$ that is the underlying communication graph, and consider a (possibly different) directed graph $G'=(V,E')$. To each node $v \in V$ is provided as input the set of its neighbors that it has in $G'$ (its outgoing edges). Notice that the neighbors that $v$ has in $G$ and in $G'$ could be different. The problem $Distance_k$ consists in finding a node $w \in V$ that is at distance $k$ from a fixed node $u$ in $G'$. A node $w$ is at distance $k$ from $u$ if there is a directed path starting from $u$ that ends at $w$ and that path is the shortest one. An example of input instance of $Distance_k$ is shown in Figure \ref{figexampledist}, where, for example, a valid result for $Distance_2(1)$ is $7$.

\begin{theorem}
	\label{thmbigmsg}
	There are instances of $Distance_k$ that can be solved in $O(k)$ rounds using messages of size $B=O(\log n)$, that, if we want to solve in $k-1$ rounds, we need to use messages of size $B = \Omega(\frac{n}{k^3})$.
\end{theorem}

In the following we will prove Theorem \ref{thmbigmsg}. Consider instances of $Distance_k$ where the out-degree of $G'$ is at most $1$. In this case, in the \CONGEST\/ model, using messages of size $O(\log n)$ bits, this problem can be solved in $O(k \cdot D(G) )$ rounds, where $D(G)$ is the diameter of the communication graph, using Algorithm \ref{algodistk}.

\begin{algorithm}[h]
	\caption{$Distance_k(u)$}
	\label{algodistk}
	\begin{algorithmic}[1] 
		\If{ my Id is $u$ }
			\State my distance is $0$
		\Else
			\State my distance is $+\infty$
		\EndIf	
		\For{$i\gets 0, k-1$}
			\If{ my distance is $i$ }
				\State broadcast in $G$ the Id of my neighbor in $G'$
			\EndIf
			\If{ I receive a broadcast containing my Id $\land$ my distance is $+\infty$ }
				\State my distance is $i+1$
			\EndIf
		\EndFor
		\If{ my distance is $k$ }
			\State broadcast my Id
		\EndIf
	\end{algorithmic}
\end{algorithm}

The idea is the following: once a node knows that it is at distance $i$, it can perform a BFS on $G$ in order to broadcast its outgoing edge, that it has in $G'$, to all the other nodes. Since each broadcast requires $O(D)$ rounds, in total $O(k \cdot D)$ rounds are required. Notice that, if the diameter of $G$ is constant, the complexity becomes $O(k)$. 

We now show that, in order to solve the problem in $k-1$ rounds, we need messages of size at least $\Omega(\frac{n}{k^3})$. Notice that, by choosing $k = \log n$, we obtain the following: there are instances of $Distance_k$ that can be solved in $O(\log n)$ rounds using messages of size $O(\log n)$, while in order to solve the problem in less than $\log n$ rounds, messages of size $\Omega(\frac{n}{\log^3 n})$ are needed, i.e., there is an exponential gap in the size of the messages.

In \cite{DBLP:journals/siamcomp/NisanW93} the following problem, $Pointer_k$, is defined. There are two players, Alice and Bob, and to each of them is provided a list of $n$ pointers, each pointing to the list of the other player. Their task is to follow these pointers, starting from some fixed pointer of Alice, and then find the $k$-th pointer. It is known that the communication complexity of this problem is $O(k \log n)$ bits: it is enough to send the $i$-th pointer at the $i$-th round, for a total of $k$ rounds, where at each round $\log n$ bits are sent. They show that, in order to solve this problem in $k-1$ rounds, even if Las Vegas algorithms are allowed, the communication complexity of this problem becomes $\Omega(\frac{n}{k^2})$ bits. Notice that for $k = \log n$ there is an exponential gap between the two complexities.

\begin{figure}[t]
	\centering
	\includegraphics[width=0.55\textwidth]{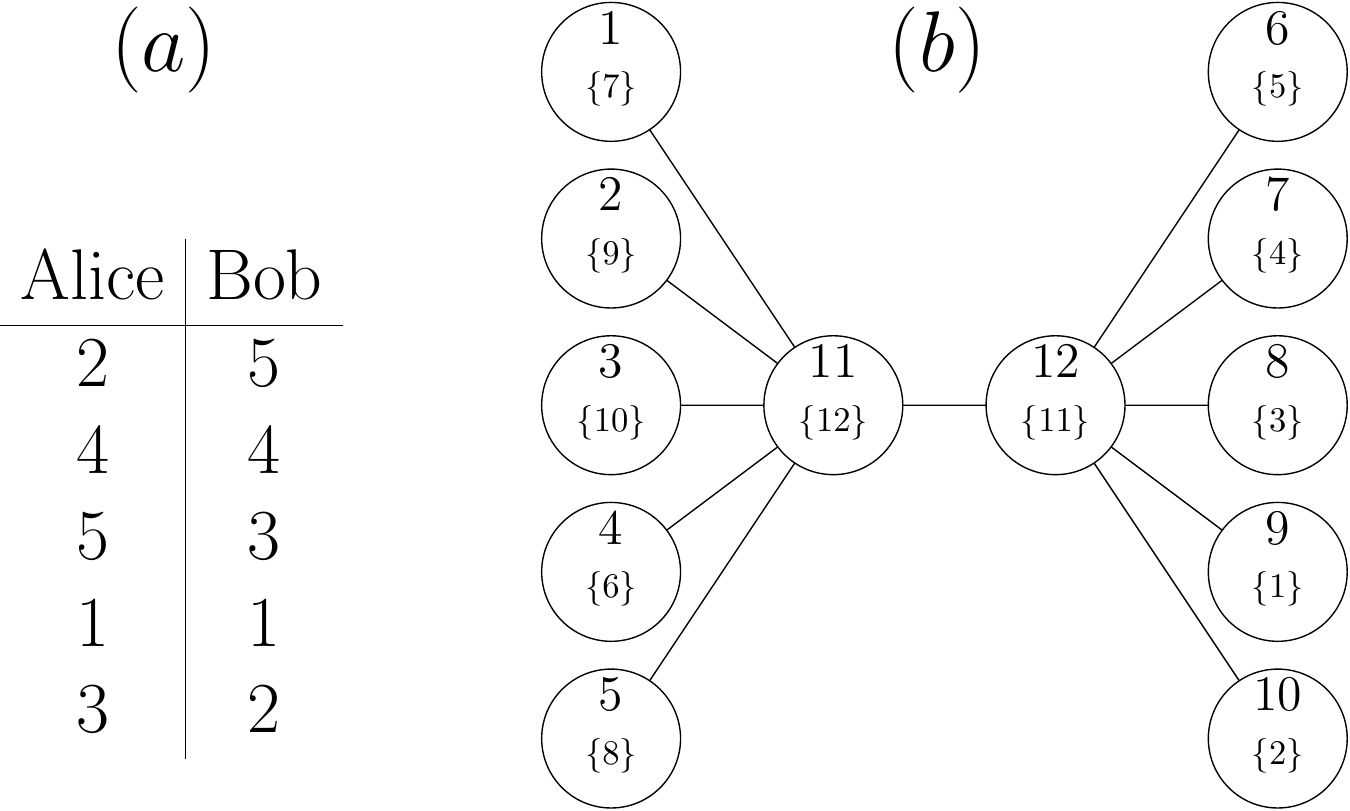}
	\caption{(a) shows the $Pointer_k$ instance, while (b) shows its $Distance_k$ version.}
	\label{reduction}
\end{figure}

We can reduce an instance of the problem $Pointer_k$ on $p$ pointers to an instance of $Distance_k$ with $p+2$ nodes in the following way. Construct a graph $G=(V,E)$ where the nodes $V$ are partitioned in two groups $L$ and $R$ of equal size, plus two special nodes $l$ and $r$ (see Figure \ref{reduction}). The nodes of each group are connected to a single special node and there is an edge connecting the two special nodes $l$ and $r$. To each node of $L$ is assigned a different ID from the set $\{ 1,\ldots,|L|\}$ and to each node of $R$ is assigned a different ID from the set $\{|L|+1,\ldots,|L|+|R|\}$. We assign to node $l$  the ID $|V|-1$ and to node $r$ the ID $|V|$. The edges of $G'=(V,E')$ are defined in the following way. If the $i$-th pointer of Alice points to the $j$-th pointer of Bob, the node with ID $i$ has an outgoing edge to the node with ID $|L|+j$. On the other hand, if the $i$-th pointer of Bob points to the $j$-th pointer of Alice, the node with ID $|L|+i$ has an outgoing edge to the node with ID $j$. Also, nodes $|V|-1$ and $|V|$ have an edge between themselves. An example of reduction is shown in Figure \ref{reduction}.
Notice that Alice (resp. Bob) can construct the input of the nodes of $L$ (resp. $R$) without communicating with Bob (resp. Alice). Assume that there is an algorithm for the \CONGESTB\/ model that solves the problem in $k-1$ rounds. Alice and Bob can simulate this algorithm by exchanging the $B$ bits that are transmitted between the two special nodes at each round. Thus, a lower bound for $Pointer_k$ holds for $Distance_k$ as well. 
\begin{figure}[h]
	\centering
	\input{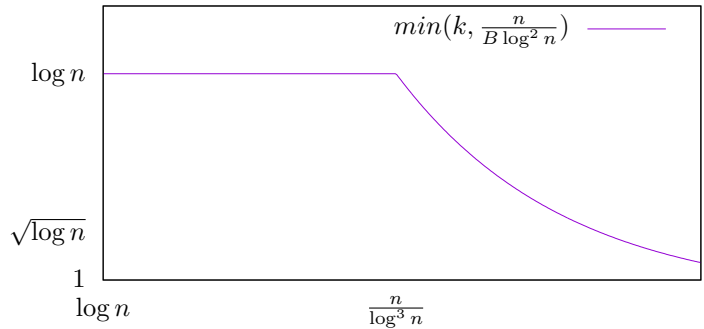}
	\caption{Round complexity as a function of B}
\end{figure}

In the \CONGESTB\/ model, we are allowed to send $B$ bits on each edge at each round. If we want to solve $Distance_k$ in less than $k$ rounds, we need to transfer $\Omega(\frac{n}{k^2})$ bits, thus $\Omega(\frac{n}{B k^2})$ rounds are required. In other words, we need to satisfy $\frac{n}{B k^2} < k$, that gives $B = \Omega(\frac{n}{k^3})$.

\section{Conclusion}
We considered the problem of understanding how the bandwidth affects the round complexity in the \CONGEST\/ model. Our results show that in some cases the round complexity perfectly scales with the bandwidth, matching existing lower bounds. On the other hand we showed that there are problems in which we need to increase exponentially the size of the messages in order to speed up the round complexity. These results suggest that it is important to analyze algorithms in the \CONGEST\/ model by considering a wider spectrum of bandwidths. It would be interesting to study this tradeoff between bandwidth and round complexity for other problems present in the literature.

\bigskip

\noindent\textbf{Acknowledgements:} I would like to thank my supervisor, Pierre Fraigniaud, for fruitful ideas and comments about the topic of this paper.

\bibliographystyle{plain}
\bibliography{Bibliography}

\end{document}